\documentclass[manuscript]{aastex}
\usepackage{longtable}

\shorttitle{SOM to stellar spectral classifications}

\begin{document}


\title{Application of Self-Organizing Map to stellar spectral classifications}


\author{Bazarghan Mahdi\altaffilmark{1}}
\affil{Department of Physics, Zanjan University, Zanjan 313, Iran
}
\altaffiltext{1}{Department of Physics, Zanjan University, Zanjan 313, Iran} 

\email{bazargan@znu.ac.ir}

\begin{abstract}
We present an automatic, fast, accurate and robust method of classifying 
astronomical objects. The Self Organizing Map (SOM) as an
unsupervised Artificial Neural Network (ANN) algorithm is used for
classification of stellar spectra of stars. The SOM is used to make
clusters of different spectral classes of Jacoby, Hunter and
Christian (JHC) library.
This ANN technique needs no training examples and the stellar
spectral data sets are directly fed to the network for the classification. The
JHC library contains 161 spectra out of which, 158 spectra are
selected for the classification. These 158 spectra are input vectors
to the network and mapped into a two dimensional output grid. The input
vectors close to each other are mapped into the same or neighboring neurons
in the output space. So, the similar objects are making clusters in the
output map and making it easy to analyze high dimensional data.

After running the SOM algorithm on 158 stellar spectra, with 2799 data
points each, the output map is analyzed and found that, there are 7 clusters
in the output map corresponding to O to M stellar type. But,
there are 12 misclassifications out of 158 and all of them are
misclassified into the neighborhood of correct clusters which gives a success rate of about 92.4\%.
\end{abstract}





\section{Introduction}

Artificial neural networks are now becoming a more popular tools for
handling astronomical data. It is very important to have some
automatic means of analyzing large databases like the 
surveys and upcoming space missions which would release terabytes of data to
the community. That is why, ANNs are widely used as an automatic tools for
analyzing astronomical data. Some of the previous attempts on
ANNs in astronomy are: \citet{b7,b9} ,\citet{b18} ,\citet{b19}
,\citet{b14,b15} , \citet{b10}, \citet{Allende}, \citet{mahdiNano}
,\cite{mahdiELODIE}, \cite{mahdiSDSS}, \cite{Wyrzykowski}.

Artificial neural network techniques used in the field of astronomy
have been mostly supervised algorithms. Here we introduce an
application of unsupervised technique to identify different classes
of objects and try to make a cluster of similar type of stars using
this technique. 

The \citet{JHC} library is used for classification using SOM algorithm as
unsupervised ANN. This algorithm configures output into a topological
presentation of the original multi-dimensional data, producing a SOM in which 
input vectors with similar features are mapped to the same map unit or nearby
units. In this way at the output map we will have the clusters of similar
objects at different positions of the map. The expected clusters in this case
are the stellar spectral type ranging from O to M.

The Self- Organizing Map algorithm is explained in section 2 . In section 3 we
describe the input data and their preprocessing. The result of classification
and discussion are presented in sections 4 and 5 respectively.

\section{Self-Organizing Map}
Self-Organizing Map neural net developed by 
\citet{Kohonen1981a, Kohonen1981b, Kohonen1981c, Kohonen1981d, Kohonen1982a, Kohonen1982b} is important from many point of view as it pays
attention to spatial order unlike the traditional neural models and
that is why this technique acquired a very special position in
neural network theory. The following section is the description of
this technique in details.

Neural network learning is not restricted to supervised learning,
wherein training pairs are provided, that is, input and target output
pairs. A second major type of learning for neural network is
unsupervised learning, in which the net seeks to find patterns or
regularities in the input data. Self-Organizing Map (SOM), developed
by Kohonen, groups the input data clusters, a common use for
unsupervised learning. The adaptive resonance theory networks, are
also clustering type networks.

\begin{figure}[h]
 \begin{center}
   \includegraphics[scale=0.35, angle=0]{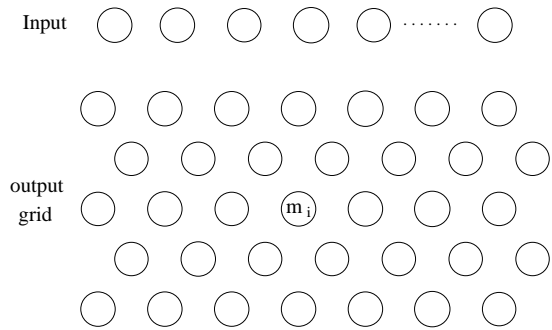}
 \end{center}
 \caption[Self-Organizing Map architecture]{Self-Organizing Map
   architecture, $m_{i}$ is the assigned weight vector of the unit on the map}
 \label{SOMnetwork}
\end{figure}
Self-organizing map consists of two layers: A one dimensional input
layer and a two dimensional competitive layer, organized as a 2D
grid of units as shown in the Figure \ref{SOMnetwork}. 
SOM lattice structure composed of $N \times N$ neural units.
This layer can neither be called hidden nor an output layer. Each input is
connected to 
all output neurons in the map. A weight vector with the same dimensionality as
the input vectors is attached to every neuron in the map.

The learning algorithm for the SOM accomplishes two important
things:
\begin{enumerate}
\item{Clustering the input data}
\item{Spatial ordering of the map so that input patterns tend to produce a
response in units that are close to each other in the grid.}
\end{enumerate}

Since every unit in the competitive layer represents a cluster, the number of
clusters that 
can be formed, will be limited by the number of units in the
competitive layer.

The weight vector for a competitive layer unit in a clustering net
serves as a representative, exemplar, or codebook vector for the
input patterns, which the net has placed on that cluster. While
training, the net determines the output units that is the best match
for the current input vector; then the weight vector of the winner
is adjusted according to net's learning algorithm.

\begin{figure}[h]
 \begin{center}
   \includegraphics[scale=0.5, angle=0]{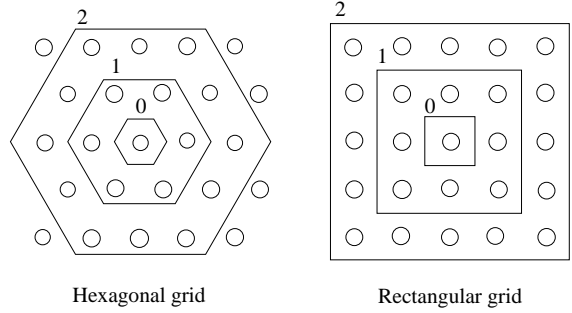}
 \end{center}
 \caption[rectangular and hexagonal lattice structures]{rectangular and hexagonal lattice structures}
 \label{lattice}
\end{figure}

The beauty of the self-organizing maps is their ability to find
regularities and correlations in the input layer and group them into
vectors without any external adjustment or prior knowledge of the
expected outcomes. Kohonen's SOM is one of the most referenced
mapping techniques because of its ability to flatten high
dimensional input into two or three dimensional data. One of the
most important attribute of Kohonen's SOM is that, it does data
compression without loss to relative distance between data points.
The neurons are connected to adjacent neuron by a neighborhood
relation, which
 decides the topology, or structure, of the map. The Figure \ref{lattice} shows
the rectangular and hexagonal lattice structures and their discrete
neighborhoods (size 0,1 and 2) of the centermost unit.


The SOM is trained iteratively. In each training step one vector $x$
from the input data set is randomly chosen and then the distance
between this vector and all the weight vectors of the map will be
calculated and this distance measurement, typically is an Euclidian
distance.

The neuron in the map whose weight vector is closest to the input
vector, $x$, is called Best-Matching Unit (BMU), here denoted by c:

\begin{equation}
\parallel x - m_{c}\parallel = \min_{i}{ \parallel x - m_{i}\parallel},
\forall i=1,2,...,N\times N
\end{equation}

After finding the BMU, the weight vector of the map are updated such
that the BMU is moved closer to the input vector in the input space.
The topological neighbors of the BMU are treated similarly. This
process of adaptation of weight vectors intern moves the BMU and its
topological neighbors towards the input sample vector $x$. 
The updating scheme aims at performing a stronger weight adaptation at
the BMU location than in its neighborhood. This update
rule in SOM for the weight vector of {\it i} is:
\begin{equation}
m_{i} (t+1) = m_{i} (t) + \alpha (t)h_{ci} (t)[x(t) - m_{i}(t)] ,
\end{equation}
where $t$ presents the time, $x(t)$ is randomly chosen an input vector
from the input data set at time $t$, $h_{ci}(t)$ is the neighborhood
kernel around the
 ${\it winner}$ unit $c$ and $\alpha(t)$ is the learning rate at time $t$. 
The function 
$h_{ci}(t)$ has a very important role; it is defined over the
lattice points. To achieve convergence it is necessary that
$h_{ci}(t)\longrightarrow 0$ as t$\longrightarrow \infty$.
Classically, a Gaussian function is used, leading to :

\begin{equation}
h_{ci}=exp - \frac{\parallel r_{c} - r_{i} \parallel^2}{2 \delta(t)^2}
\end{equation}
Here, the Euclidian norm is chosen and $r_{i}$ is the 2D location
for the $i^{th}$ neuron in the network. $\delta(t)$ speciﬁes the width
of the neighborhood during time $t$.

The learning rate $\alpha$ must always be selected much less than 1,
typically at most 0.4. Best results are usually achieved by slowly
decreasing it as training progresses, the radius of the neighborhood
around a cluster unit (winner) also decreases as the clustering
process progresses, this decreasing of the learning rate and radius
of neighborhood must be monotonic for better performance of the
network. Up to a point, large learning rate will indeed make
 training process faster. If the learning rate selected to be very large then
convergence may never occur, and in this case weight vector may
oscillate highly and it can lead to instability.

\section{The Input Data}

The \citet{JHC} library is used as a data set for the
classification and input to the SOM network. The JHC library covers
the wavelength range of 3510-7427.2 \AA~ for various O to M stellar
type and luminosity classes V, III, and I. This library contains 161 spectra of individual stars, we
selected 158 of them as an input to the SOM classifier. 
The JHC library has a resolution of 4.5 \AA~ with one sample per 1.4 \AA~.

Each spectrum of this library is having 2799 number of data points
and they are normalized to the unity before
feeding into the network. The SOM technique is actually mapping
2799 dimensional data into two dimensional output map.

The 158 input spectra are labled as the JHC's Atlas list, specifying their
spectral, subspectral and luminosity types. These lables are
require while analysing the output map.
The lables are placed beside every node (neuron) on the output map which
indicates, that specific neuron is activated by the corresponding input spectra.
Hence we can point out the location of each input spectra in the map and study
the behavior of the network in making clusters of similar spectral type.

\section{Self-Organizing Map as a classifier}
In this work the classification of the library of stellar
spectra, \cite {JHC} using unsupervised artificial neural network,
Self-Organizing Map algorithm (SOM) is presented.

The 158 spectra of the JHC library with 2799 data points each, are the input
vectors of the network and mapped into two dimensional output grid. A problem
with large 
SOMs is that when the map size is increased, the time it takes to do any 
operations on the map increases linearly. When the grid size becomes very large,
finding the best matching unit takes longer and longer. The size of the SOM
determines the resolution of the visualization it produces. On a small SOM,
lots of data samples will be projected to each SOM unit, whereas on a
large SOM the similarity relationships of the samples are more readily visible. 
The user must select the map size in advance. This may lead to many
experiments with different sized maps, trying to obtain the optimal result. 
Different map size and itterations are tried out and their performances
analysed some of which are given in Table 1. The map size of the $13 \times 13 $ with highest performance is selected for detailed analysis.

\begin{table}[h]

\centering
\begin{center}
\begin{minipage}{140mm}
\caption{List of Map sizes with number if itterations and their performances}
\begin{tabular}{ccc}
\hline Map size & No. of itterations & Performance \\
\hline
$7 \times 7 $      &    10000    &    87.34\%   \\
$9 \times 9 $      &    20000    &    84.17\%   \\
$11 \times 11 $    &    10000    &    82.27\%   \\
$13 \times 13 $    &    10000    &    92.4\%   \\
$15 \times 15 $    &    10000    &    89.87\%   \\
$17 \times 17 $    &    18000    &    81.64\%   \\
$19 \times 19 $    &    20000    &    81.01\%   \\
\hline
\end{tabular}
\end{minipage}
\end{center}
\end{table}

The 158 JHC spectral library are labeled with their true class given in the 
\citet{JHC}. The Self-Organizing Map that finds the similarity and
regularities in the input patterns, group them into clusters. Every neuron in
the grid holds a 
weight (reference) vector, the closest neuron in the map to the
input vector will win the competition. If a particular neuron
represents a given pattern, then its neighbour represent similar
patterns. This is because more than one neuron are allowed to learn
from each presentation. The learning rate for neighbours is less
than that for the winner, so their weights change less.

\begin{figure}[h]
 \begin{center}
 \includegraphics[scale=0.3,angle=0]{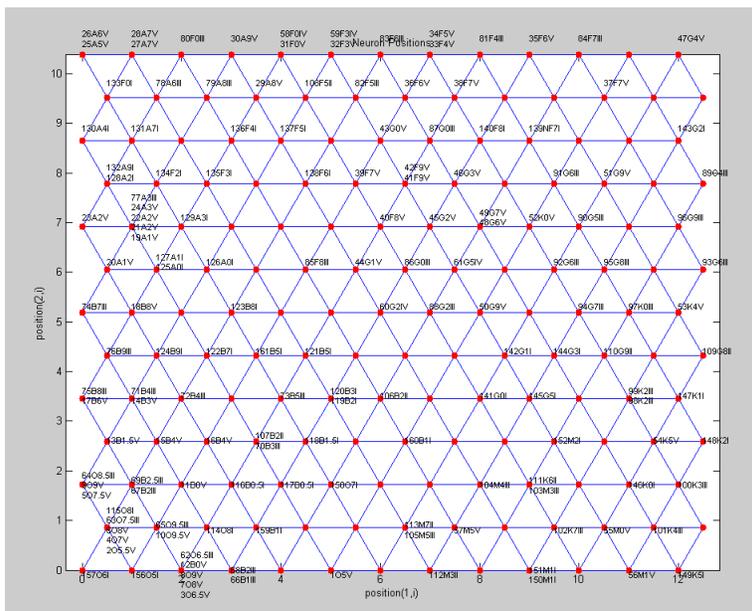}
  \end{center}
  \caption[Self-Organizing Map-1]{Kohonen output map for 158 labled JHC stellar spectra of stars as input}
  \label{SOM1}
\end{figure}

After inputing all 158 labled stellar spectra of JHC library to the network
and completion of the learning process by Neural Network, the output map is
generated. By visualizing the output map, we see that the spectrum labels
are spread out on the map by sitting on different neurons as shown in the 
Figure \ref{SOM1}. 

\begin{figure}[h]
 \begin{center}
 \includegraphics[scale=0.3,angle=0]{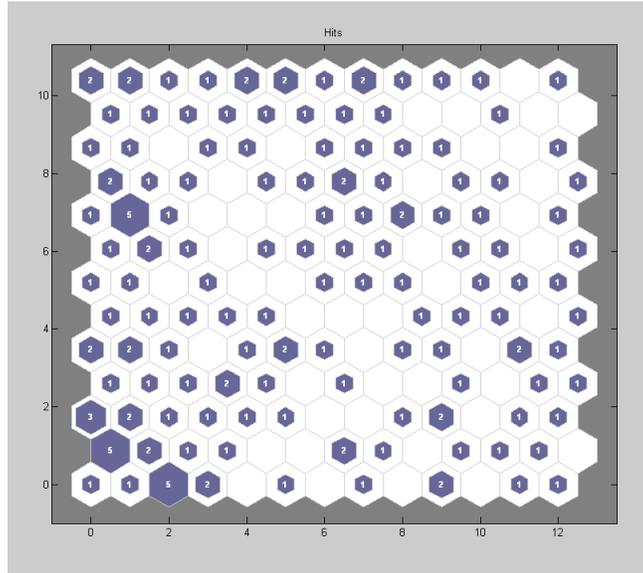}
  \end{center}
  \caption[Hit Map presentation of the output]{Number of hits indicated by
    shadowed sizes of the hexagons and also numbers printed inside each
    hexagon}
  \label{hitmap}
\end{figure}

As we see in the output map some neurons are activated with single
pattern, some with multiple patterns and some neurons have not been
activated by any spectra. The output map is also presented in another format
in Figure \ref{hitmap} which is called hit map. This map displays the number
of hits in the form of shadowed hexagons and also lables expressing digitally
the number of hits. 

This visualization method gives the
number of hits 
(number of spectra that activates the node)
for each node. The spectrum which activate single neuron 
are expected to be of the same class. 

\begin{figure}[h]
 \begin{center}
 \includegraphics[scale=0.3,angle=0]{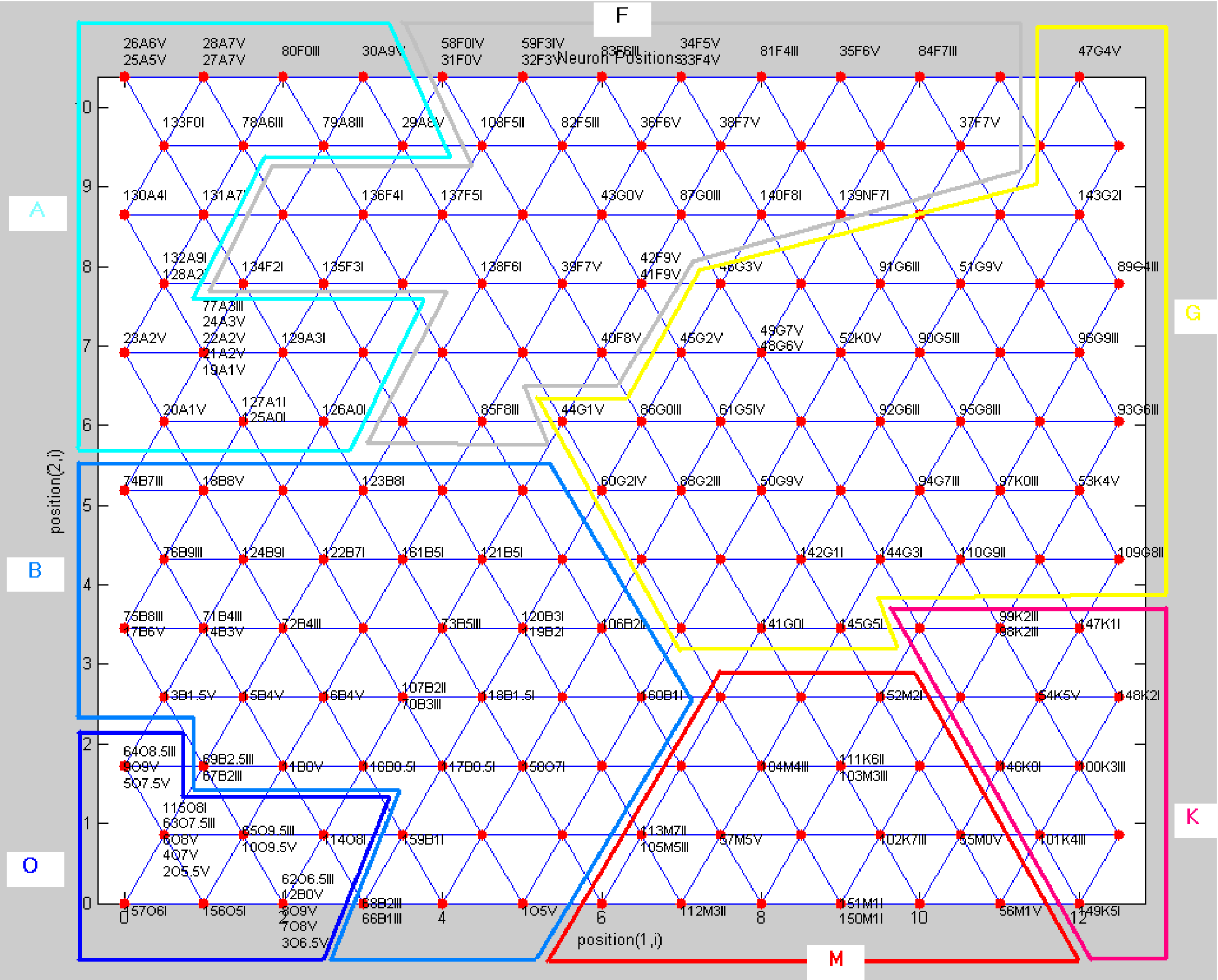}
  \end{center}
  \caption[Self-Organizing Map-2]{Kohonen output map showing labled clusters of spectral kind}
  \label{SOM2}
\end{figure}

The next step was to check the classification result and performance
of the Self-Organizing Map. We can find out the clusters and groupings by
referring to the labels in the map. The map is easily readable because of the
presence of lables in the map and their position. If any input vector out of
158 spectra activates a neuron, its lable will be printed beside that winer
neoron. 

The performance of the network can be justified by simply looking at the 
generated Kohonen output map and cheking the cluster formation throughout 
the map. The Figure \ref{SOM2} shows the output for $13 \times 13$ sized map 
with cluster and groupings of spectral types which are very nicely separated and
located in different positions of the map. The clusters for the spectral types
O - B - A - F - G - K and M are distinctly shown and separated by closed
loops. 

\begin{figure}[h]
 \begin{center}
 \includegraphics[scale=0.3,angle=0]{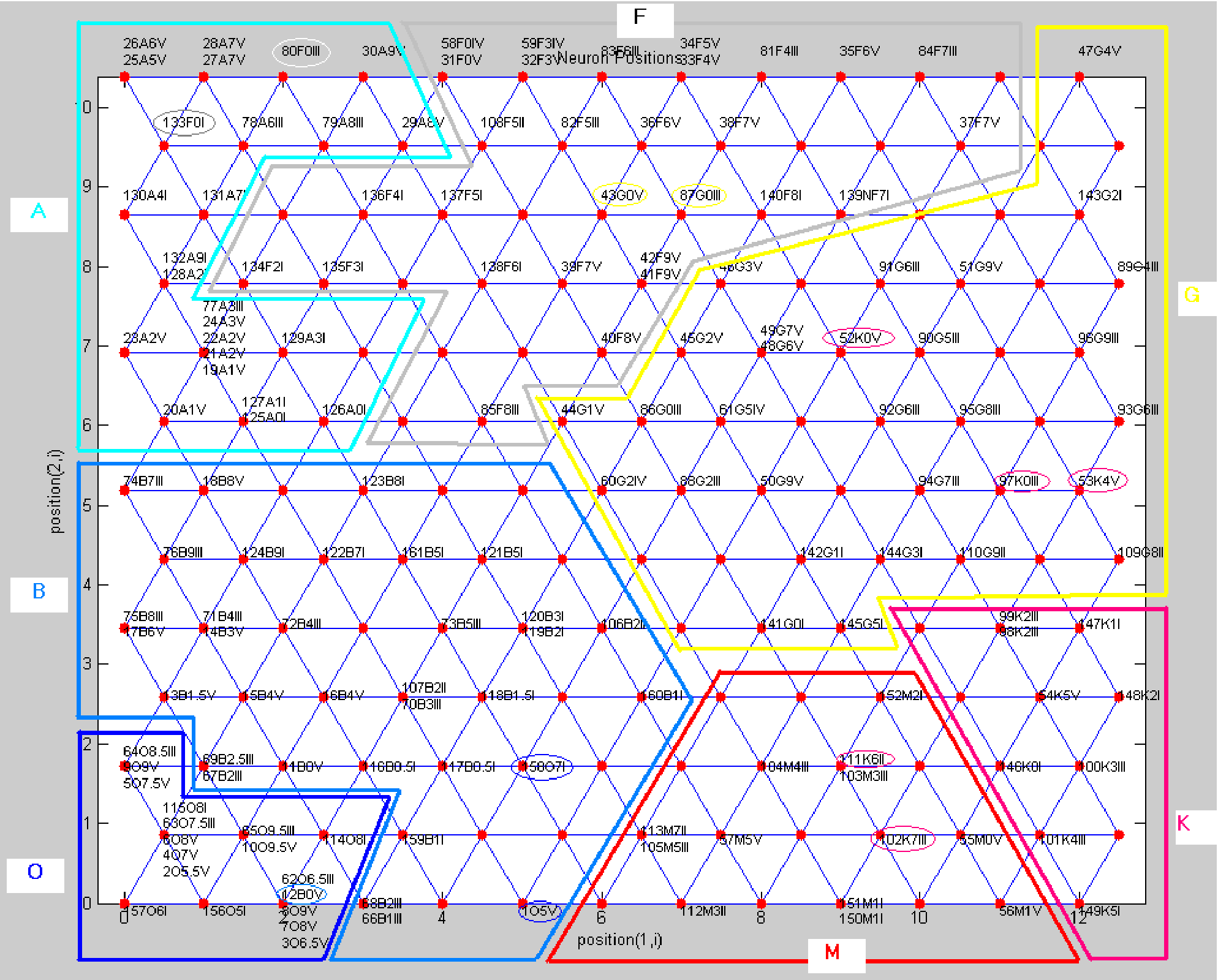}
  \end{center}
  \caption[Self-Organizing Map-3]{Kohonen output map showing misclassified
    patterns indicated with circles around the lables}
  \label{SOM3}
\end{figure}

Further investigation on classification result was done to check
 misclassifications, and it is graphically presented in
Figure \ref{SOM3}.
Misclassified patterns are shown with circles around the lables inside the 
clusters. By looking at the clusters inside the specified boundry for every
spectral type we can make out the wrongly classified and placement of the 
patterns. There are 12 misclassified patterns out of 158 spectra and 
all of them are misclassified into their neighbouring clusters as seen in Figure
\ref{SOM3} which is success rate of about 92.4\%.

The Table 2 shows the spectrum which are misclassified and their
true class types.

\begin{table}[h]

\centering
\begin{center}
\begin{minipage}{140mm}
\caption{List of spectra classified wrongly}
\begin{tabular}{ccc}
\hline Misclassified pattern & Classified as & True class \\
\hline
12B0V     &    O    &   B    \\
1O5V     &    B    &   O    \\
158O7I    &    B    &   O    \\
133F0I    &    A    &   F    \\
80F0III    &    A    &   F    \\
43G0V    &    F    &   M    \\
87G0III    &    F    &   M    \\
52K0V    &    G    &   K    \\
97K0III    &   G    &   K    \\
53K4V    &    G    &   K    \\
111K6II    &    M    &   K    \\
102K7III   &    M    &   K    \\
\hline
\end{tabular}
\end{minipage}
\end{center}
\end{table}

\section{Discussions}
As it is seen from Table 2 and Figure \ref{SOM3} misclassified
patterns belong to their neighboring clusters. For example 12B0V which is
misclassified as O spectral type, is at the neighburhood of its true
class. Likewise if we check all the misclassified patterns, we
see that they are misclassified to their neighbouring true classes. 
The JHC libraray is very detailed and it also includes the subspectral types
from 0-9. The misclassified patterns could be looked in detail to consider
the subspectral types as well. It is found that the misclassified patterns are
very close to their true class. for example, 80F0III is wrongly classified
as spectral type A, but it is close to the neighbouring class, 30A9V. This is
true with most of the misclassified patterns and it shows that the
classification accuracy can be higher than 92.4\%.

This technique and algorithm can be used for even larger unknown set
of stellar spectra. Specially where we have different astronomical objects and
need a pre-classification to identify the object types first, and then do
further analysis on each cluster. The beauty of this algorithm is that we
don't need any exemplars to train the network but it finds the similarities by
its own without any supervision and makes the clusters of similar objects.    

\section{Acknowledgments}

The author wishes to thank Lukasz Wyrzykowski for his constructive comments.

\end{document}